\documentclass[runningheads]{llncs} 
\pdfoutput=1
\usepackage[utf8]{inputenc}
\usepackage{algorithm} 
\usepackage{algpseudocode}  
\usepackage[english]{babel}
\usepackage{color}
\usepackage{colortbl,tabulary,tabularx,multirow}
\usepackage{ltxtable}
\usepackage{comment,paralist}
\usepackage{wrapfig}
\usepackage{amsmath}
\usepackage{amssymb}
\usepackage{stmaryrd}
\usepackage{braket}
\usepackage{textcomp}
\usepackage{listings}
\usepackage{theorem}
\usepackage{alltt}
\usepackage{epsfig,psfrag,subfigure}
\usepackage{graphicx}
\usepackage[colorlinks,linkcolor=blue,citecolor=blue,urlcolor=blue]{hyperref}
\usepackage{multicols}
\usepackage{balance}
\usepackage{amssymb}

\newcommand{\R}{\mathbb{R}}

\everymath{\displaystyle}

\usepackage{tikz}
\usetikzlibrary{shapes,arrows,decorations.pathmorphing,backgrounds,fit,matrix,calc}

\definecolor{todo-color}{rgb}{1,0,0}

\definecolor{ploc-color}{rgb}{0.9,0.6,0}

\definecolor{romain-color}{rgb}{1,0,1}

\definecolor{pierre-color}{rgb}{0,1,1}

\definecolor{no-color}{rgb}{1,0.2,0.2}

\renewcommand{\c}{\ensuremath{\operatorname{c}\!}}

\newcommand{\benum}{\begin{enumerate}}
\newcommand{\eenum}{\end{enumerate}}
\newcommand{\nc}{\newcommand}
\newcommand{\rnc}{\renewcommand}
\newcommand{\bvb}{\begin{verbatim}}
\newcommand{\evb}{\end{verbatim}}

\nc{\bq}{\textbf}
\nc{\m}{\textrm}
\nc{\bb}{\mathbb}
\nc{\til}{\texttildelow}
\nc{\be}{\begin{equation}}
\nc{\ee}{\end{equation}}
\nc{\dps}{\displaystyle}
\rnc{\l}{\left(}\rnc{\r}{\right)}
\nc{\lc}{\left\{}\nc{\rc}{\right\}}
\nc{\lb}{\left[}\nc{\rb}{\right]}
\nc{\ba}[1]{\begin{array}{#1}}
\nc{\ea}{\end{array}}       
\nc{\ra}{\rightarrow}
\nc{\li}{\left |}
\nc{\ri}{\right |}
\nc{\pde}[2]{\frac{\partial #1}{\partial #2}}
\nc{\ode}[2]{\frac{d #1}{d #2}}
\nc{\odee}[3]{\frac{d^{#3} #1}{d #2^{#3}}}
\nc{\pdee}[3]{\frac{\partial^{#3} #1}{\partial #2^{#3}}}
\nc{\bn}{\begin{enumerate}}
\nc{\en}{\end{enumerate}}
\nc{\bt}{\begin{theorem}}
\nc{\et}{\end{theorem}}
\nc{\y}[1]{\lambda_{#1}}
\nc{\ninf}{{\oplus}^{-\infty}}
\nc{\pinf}{{\oplus}^{+\infty}}
\nc{\nninf}{{\otimes}^{-\infty}}
\nc{\ppinf}{{\otimes}^{+\infty}}
\nc{\ir}{\mathbb{I}\mathbb{R}}

\nc{\ep}{\mathcal{E}_{P}}
\nc{\mr}{\mathcal{M}_{r}}
\nc{\mfa}{\mathcal{M}_{f,a}}
\nc{\mfp}{\mathcal{M}_{f,p}}
\nc{\mt}{\m{T}}
\nc{\F}{\mathbb{F}}

\tikzstyle{block} = [draw,rectangle,thick,minimum height=2em,minimum width=\textwidth]
\tikzstyle{sum} = [draw,circle,inner sep=0mm,minimum size=2mm]
\tikzstyle{connector} = [->,thick]
\tikzstyle{line} = [thick]
\tikzstyle{branch} = [circle,inner sep=0pt,minimum size=1mm,fill=black,draw=black]
\tikzstyle{guide} = []
\tikzstyle{snakeline} = [connector, decorate, decoration={pre length=0.2cm,
                         post length=0.2cm, snake, amplitude=.4mm,
                         segment length=2mm},thick, magenta, ->]

\usepackage{environ}

\def\comment#1{}
\newlength{\hsbw}

\tikzstyle{mybox} = [draw, very thick, rectangle, rounded corners, inner sep=0pt, inner ysep=2pt]
\tikzstyle{fancytitle} =[fill=white, draw, rectangle, rounded corners, very thick]
\newsavebox{\GrayRoundedBox}
\NewEnviron{mytikzbox}[1]{%
  \begin{lrbox}{\GrayRoundedBox}
    \setlength{\hsbw}{\linewidth}
    \addtolength{\hsbw}{-8pt}
    \begin{minipage}[b]{\hsbw}
      \begingroup\small
      \begin{alltt}
        \BODY
      \end{alltt}
      \endgroup
    \end{minipage}
  \end{lrbox}
  \begin{flushleft}
    \vspace{-10pt}
    \begin{tikzpicture}
      \node[mybox](box){\usebox{\GrayRoundedBox}};
      \node[fancytitle, left=10pt] at (box.north east) {\tiny {#1}};
    \end{tikzpicture}
  \end{flushleft}
}

\newsavebox{\Zname}
\newenvironment{tikzboxtt}[1]{%
  \sbox\Zname{\tiny {#1}}
  \begin{lrbox}{\GrayRoundedBox}
    \setlength{\hsbw}{\linewidth}
    \addtolength{\hsbw}{-8pt}
    \begin{minipage}[b]{\hsbw}
      \begingroup\small
      \begin{alltt}
}{    \end{alltt}
      \endgroup
    \end{minipage}
  \end{lrbox}
  \begin{flushleft}
    \vspace{-1em}
    \begin{tikzpicture}
      \node[mybox](box){\usebox{\GrayRoundedBox}};
      \node[fancytitle, left=10pt] at (box.north east) {\usebox{\Zname}};
    \end{tikzpicture}
  \end{flushleft}
  \vspace{-.5em}}

\tikzset{diagram background/.style={fill=blue!5,rounded corners=0.5cm}}
\tikzstyle{block} = [rectangle, draw, fill=blue!20, 
    minimum width=4em, 
    text centered, rounded corners, minimum height=3em]
\tikzstyle{proof block} = [rectangle, draw, fill=magenta!10,
    minimum width=4.2em, 
     rounded corners, minimum height=1.5em]
\tikzstyle{frama block} = [rectangle, draw, fill=green!20, minimum width = 6em,
  minimum height = 4em, rounded corners]
\tikzstyle{library block} = [rectangle, draw, fill=gray!20, minimum width = 4em,
  minimum height = 2em, rounded corners]
\tikzstyle{invisible} = [opacity=0] 
\tikzstyle{every picture}+=[remember picture]

\theoremstyle{plain} \theorembodyfont{\upshape}

\begin{document}

\mainmatter

\title{Automated, Credible Autocoding of An Unmanned Aggressive Maneuvering Car Controller}

\titlerunning{Automated, Credible Autocoding of An Unmanned Vehicle}

\author{Timothy Wang\inst{1} \and  Éric Féron\inst{1}}

\authorrunning{ }

\institute{ Georgia Institute of Technology,
Atlanta, Georgia, USA}

\maketitle

\begin{abstract}
This article describes the potential application of a credible autocoding framework for control systems
towards a nonlinear car controller example.
The framework generates code, along with guarantees of high level
functional properties about the code that can be independently verified.
These high-level functional properties not only serves as a certificate of good system behvaior but also
can be used to guarantee the absence of runtime errors. 
In one of our previous works, we have constructed a prototype autocoder with proofs that demonstrates 
this framework in a fully automatic fashion for linear and quasi-nonlinear controllers. 
With the nonlinear car example, we propose to further extend the prototype's dataflow annotation language environment with 
with several new annotation symbols to enable the expression of general predicates and dynamical systems. 
We demonstrate manually how the new extensions to the prototype autocoder work on the car controller using the output language Matlab. 
Finally, we discuss the requirements and scalability issues of 
the automatic analysis and verification of the documented output code. 
\end{abstract}

\section{Introduction}
\label{sec:intro}
In this article, we examine the autocoding of a nonlinear car controller model along with its control semantics. 
The system is of interest because of the presence of fully nonlinear dynamics and a nonlinear non-quadratic invariant. 
We discuss the impact of these nonlinearities on the autocoding and the proof-checking. 
Other issues of interest include the presence of transcendental functions and their impact on the correctness of the control-theoretic proofs. 

This work follows~\cite{fcsm10} that first described the idea of leveraging control domain proofs
for code analysis and~\cite{wjfarxiv11}~\cite{heber}~\cite{wjfarxiv13}, which are later efforts in creating an automated 
framework that can perform such leveraging on real case studies. 

The article is organized as follows: we first give a brief summary of the automated framework in section~\ref{sec:problem};
we then proceed to describe the model of the car and its control-theoretic properties in section~\ref{sec:car_model}. 
Sections~\ref{sec:control_semantics}~\ref{sec:autocoding} describe informally, the expression and translation of the control-theoretic properties of 
the nonlinear car controller from the Simulink model to Matlab code. 
Finally, in section~\ref{sec:autoverif}, we describe the limitations encountered in the automatic verification of 
the fully-nonlinear case.

\section{Autocoding with Control Semantics} 
\label{sec:problem}
First we give a brief summary about the automated framework that leverages control system knowledge for 
the formal analysis of code and the prototype that we built to demonstrate this framework in a fully automated fashion. 
For a good introduction on the control stability proofs and code analysis, one can refer to ~\cite{fcms10}. 
For more details on the automated framework and the prototype implementation, one can refer to ~\cite{wjfarxiv13}. 

We define, informally-speaking, 
control semantics as any innate properties of the control system that can be expressed either as a predicate or as 
a state-transition system. 
This can included, but is not limited to, proofs of stability or absolute stability, performance measures like margins and overshoot, 
and elements that are used in the construction of the proofs such as the model of the system being controlled i.e. the plant, or other dynamical systems 
used in the construction of the proof or the controller. 
The framework of autocoding with control semantics 
is consisted of the following: a synchronous modeling language environment extended with synchronous
observers to express the control semantics on the model level; an autocoder extended with
the capability of transforming the observers into useful annotations for the code output;
a proof-checking analyzer that can automatically verify the annotated code.
\begin{figure}
	\includegraphics[scale=0.38]{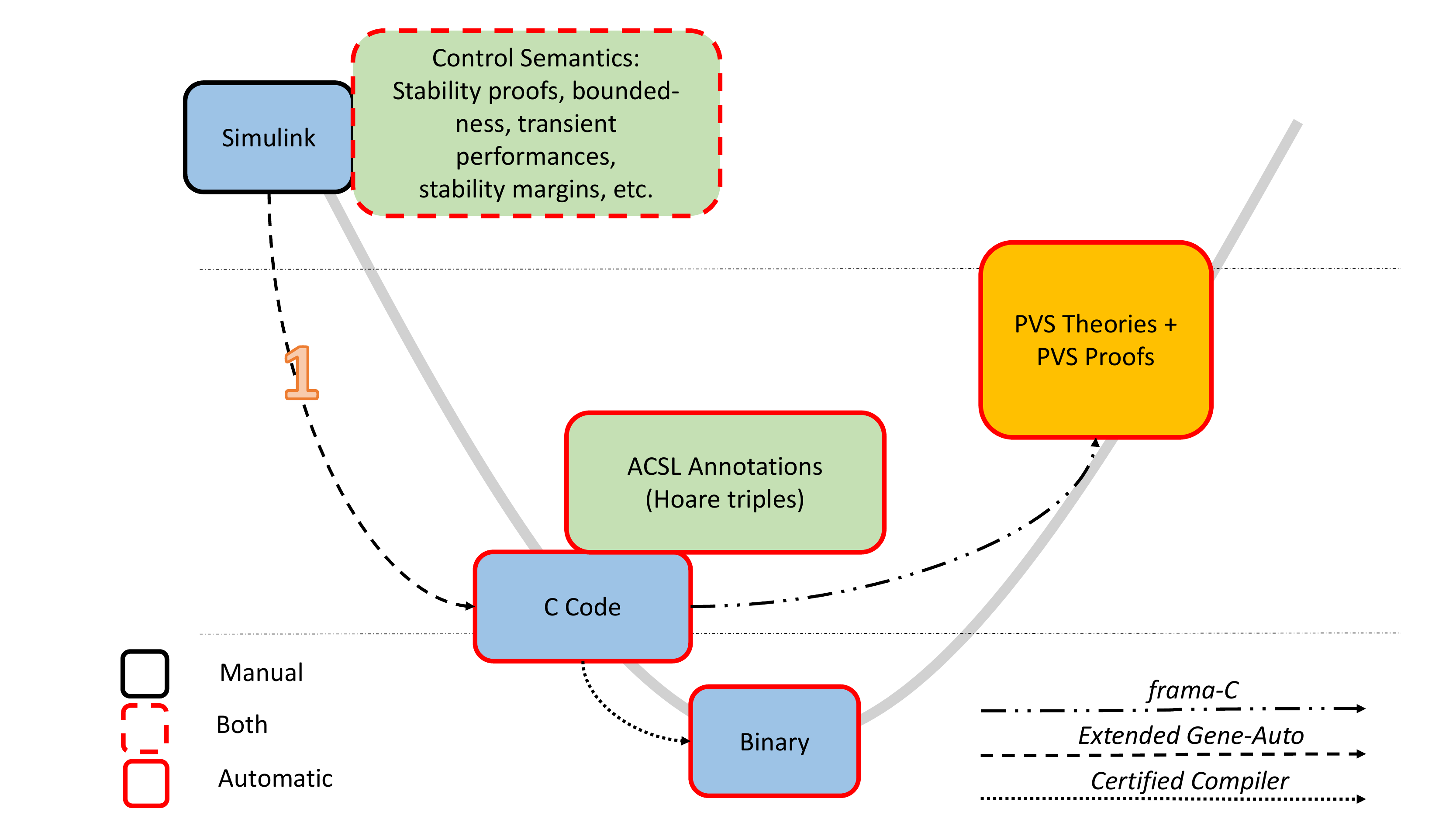}
	\caption{Autocoding with Control Semantics from Simulink to C \cite{wjfarxiv13}}
	\label{ver}
\end{figure}
As shown in figure \ref{ver}, the input language of the prototype was chosen to be Simulink, and the output language is C. 
These two languages were chosen due to their ubiquitous presence in the industry as we want the prototype to be  
a practical demonstration of the framework that we just discussed. 

The prototype autocoder in~\cite{wjfarxiv13} has been built to handle linear controllers and quasi-nonlinear cases that are 
linear systems with bounded nonlinearities in a feedback interconnection configuration.  
The prototype is based on the Simulink-to-C autocoder Gene-Auto~\cite{nassimaformal09}. 
For linear or quasi-nonlinear controllers, the leveraging of the control system knowledge 
comes from the fact that the proof of stability for these systems forms a quadratic invariant of the following form 
\be
\dps x^{\m{T}} P x \leq 1
\label{quadratic}
\ee, where $x \in \R^{n}$ denotes an array of states and $P$ is a $n \times n$ symmetric matrix. 
The quadratic formula in (\ref{quadratic}) can be used to show that the variables of the output code that correspond to the states in $x$ are bounded. 
Furthermore, any variables in the output code that are related linearly to the states in $x$ are also bounded.
Hence the proof of stability, which results in the matrix $P$, can be used both as a certificate of good behavior for a control system and 
to show the absence of runtime errors. 
For nonlinear controllers, we can apply the same type of leveraging, with caveat that the proof of stability
is does not usually result in a simple symmetric matrix $P$. 
In the next section, we present a car example that is fully nonlinear and its proof of stability does not yield a quadratic invariant.

\section{Car Controller Example} 
\label{sec:car_model}
We choose the running example to be the sliding mode based nonlinear car controller from \cite{velenis01}.
Here we give a detailed overview of its specifications. 
For the sake of consistency and simplicity, we have chosen to reuse the notations from \cite{velenis01}. 
There are two main reasons why we picked this car controller: 
\begin{enumerate}
\item Unlike the previous examples that we have studied in the context of autocoding with control semantics, 
this example is fully nonlinear. 
By fully nonlinear, we mean that the car controller cannot be reformulated as a linear system with bounded nonlinearities in a 
feedback interconnection. 
\item The car controller is constructed using an non-quadratic Lyapunov function. The non-quadratic Lyapunov function 
results in a non-quadratic nonlinear invariant, which our prototype autocoder could not handle.  
\end{enumerate}

\subsection{Single Track Model} 
\begin{figure}
\centering
\includegraphics[scale=0.37]{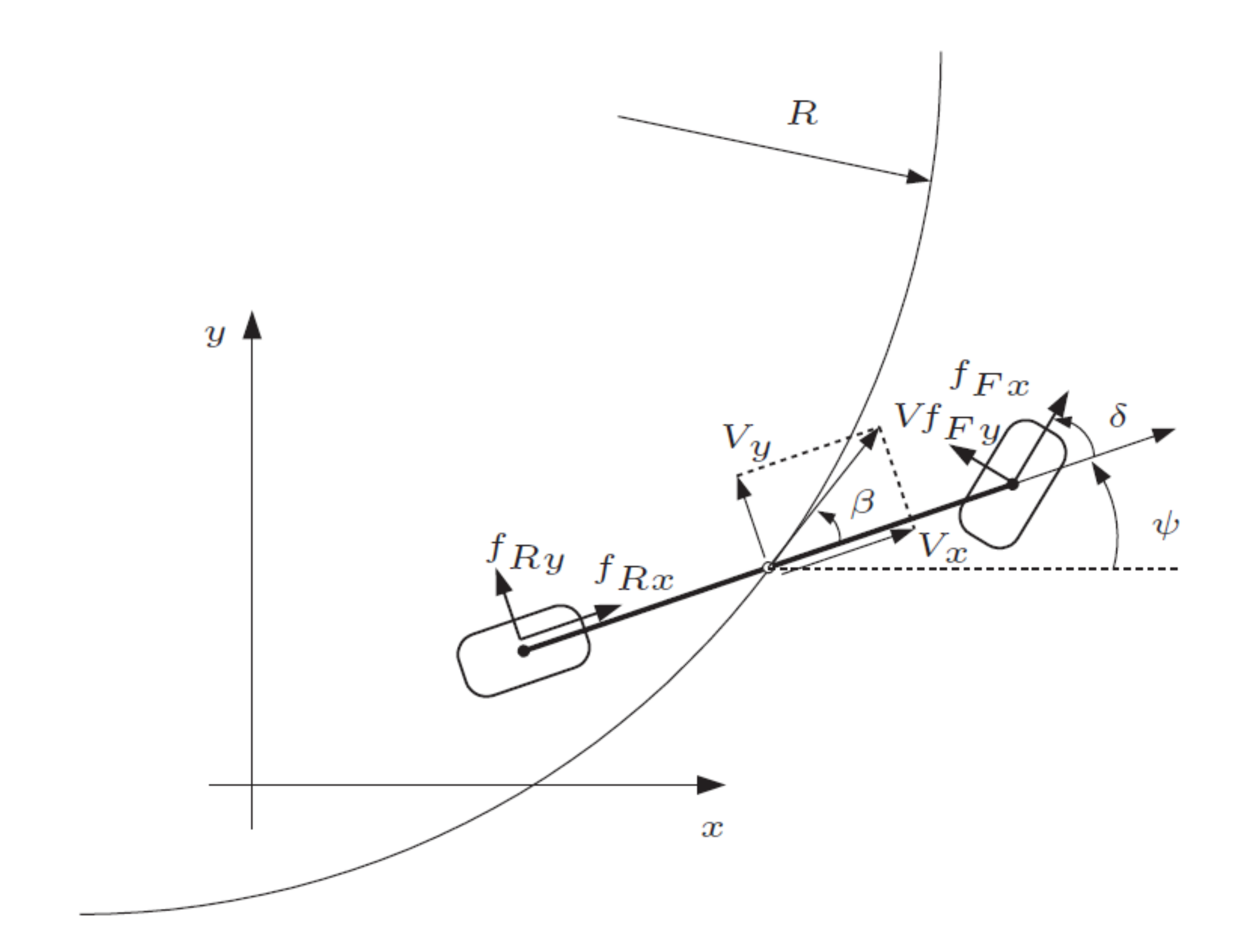}
\caption{Single-track Model of the Car \cite{velenis01}}
\label{tire_model}
\end{figure}

In figure \ref{tire_model}, we have the single-track model of the car. 
The wheels have a radius of $r$ and angular velocities of $\omega_{i}$, $i=F,R$, in which $F,R$ denote respectively the front and rear locations.   
The controller stabilizes the single-track model of a car about a steady-state cornering equilibrium. 
In a steady-state cornering equilibrium, the car undergoes a constant radius turn in which its speed $V$, path angle $\beta$, steering angle $\delta$,  and angular velocity $\dot{\psi}$ are all constant.
In additions, the longitudinal slips $s_{i}$, $i=F,R$ are also constant. 
The longitudinal slip is a measure of the amount of "sliding"
due to a difference between the tangential speed of a wheel ($\omega r$) and
 the velocity $V$ of the car projected onto the longitudinal axis of the wheel. 
From the model in figure (\ref{tire_model}), the longitudinal slips are
\be\ba{l} 
\dps s_{F}=\frac{Vcos\l \beta - \delta \r + \dot{\psi} l_{f} sin\l \delta \r - \omega_{F}  r}{\omega_{F} r }  \\
\dps s_{R}=\frac{V cos \l \beta \r - \omega_{R} r } {\omega_{R} r} 
\ea
\label{slips}
\ee. 

The car dynamics is represented by the nonlinear state-space system
\be
\dps \dot{x} = f \l x, u\r
\label{nonlinear_plant}
\ee
with the state vector $x=\lb \ba{ccc} V & \beta & \dot{\psi} \ea\rb^{\m{T}}$, the input 
vector $u=\lb \ba{cc} s_{F} & s_{R} \ea \rb^{\m{T}}$, and the vector field $f : \R^{5} \ra \R^{3}$. 
The precise structure of the function $f$ is not necessary for the discussion in this paper hence is omitted. 
We assume that $f$ does satisfy all the well-posed conditions to guarantee the existence, uniqueness and smooothness of solutions to
the differential equation in (\ref{nonlinear_plant}). 
The linear plant model is obtained by 
linearizing the nonlinear model in (\ref{nonlinear_plant}) 
about the cornering equilibrium $x_{ss}=\lb \ba{ccc} V_{ss} & \beta_{ss} & \dot{\psi}_{ss} \ea\rb^{\m{T}}$ and 
$u_{ss}=\lb \ba{cc} s_{Fss} & s_{Rss} \ea \rb^{\m{T}}$.
By definition, we have that $f(x_{ss},u_{ss})=0$. 
\begin{figure} 
\centering
\includegraphics[scale=0.61]{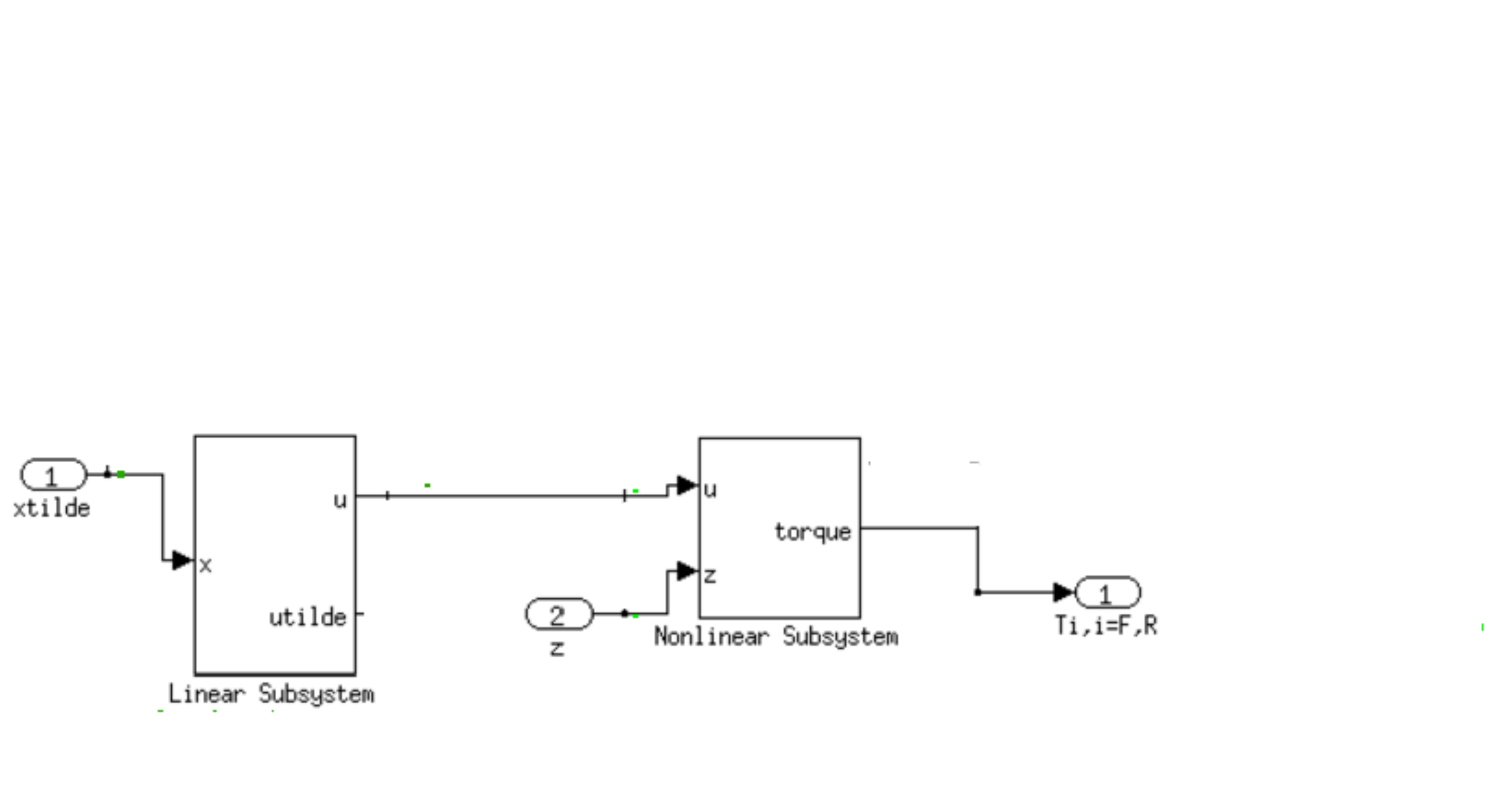}
\caption{Architecture of the Controller}
\label{arc}
\end{figure} 
The controller model can be divided into two subsystems as shown in figure \ref{arc}.

\subsection{Linear Subsystem}
This subsystem is consisted a LQR controller that drives the trajectories of the linearized model to the equilibrium point $x_{ss}$. 
The input to this controller is $\tilde{x} = x - x_{ss}$.  
We have the controller
\be
\dps \tilde{u}= -K \tilde{x}
\label{linear}
\ee
and the controller output $u=\tilde{u} + u_{ss}$. 
The $K$ is computed using the algebraic Riccati that is associated with the LOR problem. 
\subsection{Sliding Mode Subsystem}  
The output $u$ from the linear subsystem, being the longitudinal slips, cannot be controlled directly. 
Instead, we have a sliding mode controller that controls the angular velocities of the wheels by applying torques $T_{i}$, $i=F,R$. 
From (\ref{slips}), we have the following mappings $\phi_{i}$, $i=F,R$ from given longitudinal slips $u$ to the angular velocities of the wheels. 
\be
\dps \tilde{\omega} = \phi \l x , u \r =\lb \ba{l} \dps \frac{Vcos\l \beta - \delta \r + \dot{\psi} l_{f} sin\l \delta \r}{\l 1+s_F\r r}\\ \dps \frac{V cos \l \beta \r }{\l 1+s_R\r r} \ea \rb.
\label{slipwheel}
\ee
The goal is to drive the angular velocities $\omega$ of the wheels to mach the commanded 
angular velocities $\tilde{\omega}$ so the longitudinal slips are equal to $u$. We want $\omega  - \phi(x,u) =0$, which defines the manifold
\be
\dps z_{i} = \omega_{i} - \phi_{i} \l \ldots \r =0,\m{ }i=F,R. 
\label{sliding_manifold}
\ee
Let $fx_{i} $ being the longitudinal frictional forces on the wheels and let $I_{w}$ be the moment of inertia of the wheels. 
The dynamics of the rotations of the wheels is
\be
\dps  I_{w} \dot{\omega}_{i} =T_{i} - fx_{i} r , \rm{  }i=F,R.  
\label{torque_dyn}
\ee
 
With (\ref{sliding_manifold}) and (\ref{torque_dyn}), ollowing auxiliary dynamic system
\be
\dps \dot{z}_{i} = \frac{1}{I_{w}} \l T_{i} - fx_{i}r - I_{w}  \frac{\partial \phi_{i} } {\partial x} f(x,u) \r 
\label{auxiliary} 
\ee
The controller hence was picked to be
\be
\dps T_{i}=   fx_{i} r + I_{w}  \frac{\partial \phi } {\partial x} f(x,u) -  sat \l z_{i} \r, \rm{  } i=F,R. 
\label{controller}
\ee
The first part of the controller $ fx r+ I_{w}  \frac{\partial \phi } {\partial x} f(x,u)$ ensures $\dot{z}=0$ when $z=0$.
The second part of the controller $-sat \l z \r$ drives the system trajectories towards the manifold $z=0$.

\section{Annotating the Model with Control Semantics}
\label{sec:control_semantics}

We have already for linear controllers, constructed two annotation blocks tailored for linear systems. 
One is the plant block, which has the semantics of a linear state-space system. 
The other is an ellipsoid
block, which is a synchronous observer with the semantics of a quadratic inequality i.e. the block takes in an input 
of $x$ and returns if $x^{\m{T}} P x \leq 1$ is true or false. 
For fully-nonlinear controllers, we require a less restriction on the structure of the plant or the synchronous observer. 
Hence propose extending the current prototype's input Simulink environment with two new
annotation blocks. 

One is a general synchronous observer block that accepts any logic formula that can be expressed by the underlying language. 
The other one is a generalized plant block. Likewise, there is no restriction on the type of state-transition 
system that can define inside of the block. 
For a practical prototype demonstration, we can pick the underlying language to be that of the subset of Simulink supported
by Gene-Auto. 

For the example in this article, the controller model does not contain any signals with memory.
For control systems without dynamics, however unrealistic that might be, a closed-loop analysis
with the plant model becomes a necessary step before we can express any interesting high-level functional properties such 
as Lyapunov stability. 
Hence for the car controller, it is required to embed both the linearized plant model 
and the auxiliary dynamics in (\ref{auxiliary}) into the model to support the eventual
verification of the stability proofs on the code. 

We can express the linearized model by using the existing plant block. 
This is an annotation block that has already been implemented in the prototype autocoder.
The auxiliary dynamics requires the proposed generalized plant block. 
The stability proof for the linear portion of the controller can be expressed using the existing ellipsoid block.
The proof of the stability of the sliding-mode portion of the controller require the new generalized synchronous observer. 

\begin{figure}[htp]
\centering
\includegraphics[scale=0.61]{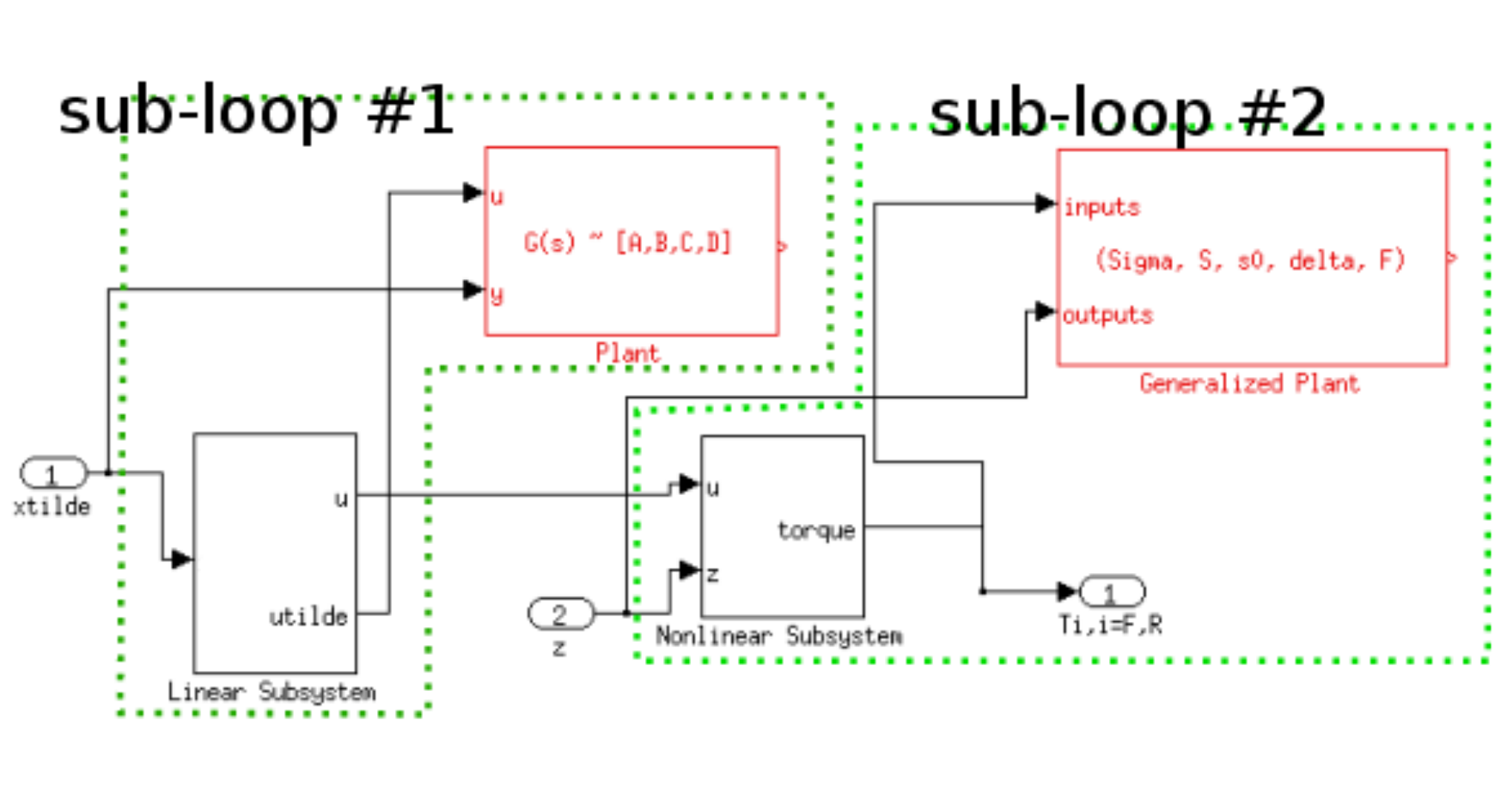}
\caption{Annotated Car Model}
\label{annot_model}
\end{figure}

For the insertion of the plant semantics and the stability proofs, we can use the same type of method as discussed 
in our previous work~\cite{wjfarxiv11}. 
For example, to insert the linear plant block into the diagram, one can
simply connect the output signal labelled $utilde$ from the linear subsystem 
into the input port on the plant block that is labelled $inputs$ or $u$
and then connect the input signals of the linear subsystem into the input port on 
the plant block that is labelled $outputs$ or $y$. 
Similarly we can connect the auxiliary dynamics with the nonlinear subsystem using the proposed generalized plant block. 
The annotated Simulink model is shown in figure \ref{annot_model}. 
For loop \#1, since it is a stable linear system, we can compute a matrix $P$ that forms a quadratic invariant. 
For loop \#2, we have the non-quadratic invariant $z\l \omega, x, u\r^{\m{T}} z\l \omega,x,u\r \leq 1$ and $z$ is a nonlinear function of the states $\omega, x$ and inputs $u$ 
as defined in (\ref{sliding_manifold}). 
Note the annotation blocks for expressing the stability proofs are not displayed in figure \ref{annot_model} since it is not always required for them to be explicitly inserted into the model. 
For the linear cases, the prototype autocoder can analyze the model and compute a stability proof. 
The insertion of the stability proofs as an invariant is done internally in the autocoding process. 
For the fully-nonlinear case, this process cannot be fully-automated for the most part and requires the end user to manually insert the proof stability using a general synchronous observer block. 

Since we have not finalized the grammar and semantics of the proposed new annotation blocks, we choose to skip those details in this article. 
Unlike the examples that we have looked at before, the two invariants for the two loops of the system require 
two completely different set of analysis and autocoding methods. 
One is linear with a quadratic invariant which is fully handled by the current version of the prototype autocoder. 
The second loop has an nonlinear non-quadratic invariant, which is currently partially supported by the prototype. 
We focus on the nonlinear part in the next section in a manual demonstration of the analysis and autocoding of the non-quadratic invariant.

\section{Example of the Autocoded Output} 
\label{sec:autocoding}
For the reason of brevity, in this article, 
we choose to use Matlab to represent the generated code and a Matlab annotation language that is similar to
ACSL~\cite{baudin:acsl} for the example annotations. 
We also conceal a large portion of the controller's
code inside function calls and only display a few examples annotations
for demonstration's purpose. 
Figure \ref{annot_table} has the code and the partial of set of the annotations. 
Generally speaking, the placement of the invariants from the model onto the code depends only on the type of the invariant. 
The assumptions are inserted at the location where the variables of the invariant are assigned to for the first time. 
The inductive ones e.g. all the invariants derived from the control stability proofs are inserted at the beginning and end of the code portion of their corresponding loop. 
By loop, we meant the loop 
formed by the controller subsystem (code) and its corresponding plant model (annotations)
as shown in figure \ref{annot_model}. 
For example, the quadratic invariant $\lc xtilde'*P*xtilde <= 1\rc$ for loop\#1 
is inserted into the code at line $3$ and at line $10$ of figure \ref{annot_table}. 
They become a pre and post-condition pair for the code portion of loop\#1. 
Likewise, the non-quadratic invariant $\lc z'*z <=1 \rc$ 
is inserted into the code at line $14$ and line $32$. The code between 
$14$ and $32$ is generated from the nonlinear controller subsystem of loop\#2.  

The representation and 
placement of the plant part of the loops can be arbitrary. 
Depending on the accepted input format of the backend analyzer, we can change its form and location. 
In this manually annotated example, we choose to represent the plant models 
as assumptions inside the annotations. These assumptions, when used in conjunction with a pre-condition 
can be used to prove the inserted post-condition such as the one line $32$ of figure \ref{annot_table}. 
\begin{figure}[htp]
\centering
\begin{tikzboxtt}{Matlab+Annotations} 
1: xtilde=Input(); 
2: /*@ 
3: 	require xtilde'*P*xtilde<=1
4:	ensures [xtilde; utilde]'*Q1*[xtilde; utilde]<=1;
5: */
6: utilde=K*xtilde;
7: /*@  
8	requires  [xtilde; utilde]'*Q1*[xtilde; utilde]<=1;
9:	assumes  xtilde=A*xtilde+B*utilde; 
10:	ensures  xtilde'*P*xtilde<=1 && xtilde'*Q2*xtilde <=1; 
11:*/ 
12: u=utilde-uss; 
13:/*@ 
14:	requires  z'*z<=1
15:	requires   z+dt*(1/Iw*(friction_func(x,u)*r+Iw*dphi'*f_func(x,u)
16:		-saturate(z)-friction_func(x,u)*r - Iw*dphi'*f_func(x,u)))]'*
17:		[z+dt*(1/Iw*(friction_func(x,u)*r+Iw*dphi'*f_func(x,u)-saturate(z)
18:		-friction_func(x,u)*r - Iw*dphi'*f_func(x,u)))]<=1; 
19: */
20: z=Input(); 
21: f=f_func(x,u); 
22: dphi=dphi_func(x,u);
23: friction=friction_func(x,u);
24: torque=friction*r+Iw*dphi'*f-saturate(z); 
25: Output(torque);
26: /*@ 
27:	requires [z+dt*(1/Iw*(torque-friction_func(x,u)*r 
28:	- Iw*dphi'*f_func(x,u)))]'*[z+dt*(1/Iw*(torque
29:	-friction_func(x,u)*r - Iw*dphi'*f_func(x,u)))]<=1;
30:	assumes z=z+dt*(1/Iw*(torque-friction_func(x,u)*r 
31:	- Iw*dphi'*f_func(x,u)));
32:	ensures z'*z<=1; 
33: */
\end{tikzboxtt}
\caption{Example of Annotated Autocoded Output} 
\label{annot_table}
\end{figure}
The quadratic invariant can be propagated through the code using the usual affine transformation 
techniques until end of the linear portion of the code. This procedure results in the
second post-condition i.e. the ellipsoid $\lc xtilde'*Q2*xtilde<=1 \rc$ at line $10$ of figure \ref{annot_table}. 
This ellipsoid represents the strongest post-condition, hence it is required for the validity of
$\lc xtilde'*Q2*xtilde<=1 \rc \ra \lc xtilde'*P*xtilde<=1 \rc$ to be verified. 
This is a task that should be left to the backend analyzer. 

For the non-quadratic invariant $\lc z'*z<=1 \rc$, since 
cannot be propagated using affine transformation rules, we must employ backward propagation 
using the axiomatic rules for the simple imperative language from hoare logic~\cite{hoare:logic}. 
The same problem applies if the line of code does not represent an affine transformation. 
For example in figure \ref{annot_table}, given the assumption in line $30$, the pre-condition 
in line $27$ can be generated using the assignment rule from the post-condition on line $32$. 
This backward propagation procedure is executed 
until it is terminated right before the inserted pre-condition on line $14$. 
This results in the second pre-condition on line $15$. 
Since this is a weakest pre-condition, 
we also need to check if the inserted invariant $\lc z'*z<=1\rc$ 
implies the invariant in line $15$.

\section{Automating Proof Checking} 
\label{sec:autoverif}
For the two invariants in line $14$ and $15$ in figure \ref{annot_table}, we have $A=\lc X: V(X)\leq 1\rc$ and $B=\lc Y: W(Y) \leq 1 \rc$, where $X$ and $Y$ are lists of variables. 
The backend analyzer needs to be able to automatically check that $A \ra B$. 
Knowing that $A \ra B$ if $W(Y) - V(X) \leq 0$ and that $W(Y) - V(X)$ can be simplified to another function $U(X)$, the problem 
is reduced to determine if $U(x)>0$.
Since we already know that the function $U(x)$ is positive i.e. $U(X)>0$ is always true. 
We know this is a fact because the specification of the controller 
is extracted from the proof of good behavior i.e. the Lyapunov stability proof. 
Due to the limitations of the backend analyzer, it is necessary to 
insert the invariant $U(x)>0$ as an annotation into the code to automate the proof-checking process. 
Note that $U(X)$ does not necessary belong to a class of polynomials or any other smooth functions 
because of the saturation operator in the controller, 
Hence we cannot determine 
if $U(x)$ is positive using a single general decision procedure. 
There remains a need for a theorem prover. 

\subsection{Non-quadratic Invariant and Runtime Errors} 

From the non-quadratic invariant, we can obtain bounds on many variables of the output code. 
Note that
\be
\dps z=\omega- \phi \l x,u\r
\label{slipwheel2}
\ee and $\phi \l x ,u \r$ as defined in (\ref{slipwheel}), and the steering angle $\delta$, the radius $r$ and $l_{f}$ are all fixed. 
We also have that $x=\lb \ba{ccc} V & \beta & \dot{\psi} \ea \rb^{\m{T}}$ belongs to an ellipsoid. 
We can assume that the longitudinal slip belongs to the set $\l -1, \right . \left . \infty \rb $. 
It is only equal to $-1$ in the physically impossible case where
$V=0$, $\dot{\psi}=0$, and $\omega>0$ i.e. the car is stationary while the tires are spinning. 
The auxiliary variable $z$ is bounded by $z^{\m{T}} z \leq 1$. 
Hence we know that both $\omega$ and $\phi \l x,u\r$ will be bounded. 
This decision procedure used to obtain the bounds is by no mean a fully automated one 
since it clearly depends on certain knowledge about the physical limitations of the car. 
Otherwise, we can't make any conclusions about bounds on $\omega$ and $\phi$. 

\section{Conclusion}
We have presented an application of the autocoder with control semantics framework on a nonlinear car controller example. 
The set of improvements on our prototype autocoder resulting from the car controller can be summarized as follows:
a new generalized synchronous observer block that is not restricted to any structure with the exception of the limitations
presented by the subset of the Simulink blocks supported by Gene-Auto; likewise we also generalized the plant block to be able
to express nonlinear state-transition functions.

\section{Acknowledgements}
The authors would like to thank the Army Research Office for their support under MURI Award W911NF-11-1-0046.  

\bibliographystyle{plain}
\bibliography{complete} 

\end{document}